\documentclass[aps,prl,twocolumn,graphics,epsfig,floats]{revtex4}
\usepackage{graphicx}

\newcommand{\be}{\begin{eqnarray}}
\newcommand{\ee}{\end{eqnarray}}

\def\l{\langle}
\def\r{\rangle}

\begin{document}

\title{Theory of an Entanglement Laser }

\author{Christoph Simon$^{1}$ and Dik Bouwmeester$^{2}$}
\address {$^1$ Department of Physics, University of Oxford, Parks Road, Oxford OX1
3PU, United Kingdom\\  $^2$ Department of Physics, University of
California at Santa Barbara, CA 93106}

\date{\today}

\begin{abstract}We consider the creation of polarization entangled light from parametric down-conversion
driven by an intense pulsed pump inside a cavity. The multi-photon
states produced are close approximations to singlet states of two
very large spins. A criterion is derived to quantify the
entanglement of such states. We study the dynamics of the system
in the presence of losses and other imperfections, concluding that
the creation of strongly entangled states with photon numbers up
to a million seems achievable. \pacs{}
\end{abstract}

\maketitle

Entanglement of light has mainly been demonstrated at the
few-photon level. It is a challenging goal to produce entangled
states involving large numbers of photons, approaching the domain
of macroscopic light.  Here we propose a scheme that is based on
the non-linear optical effect of parametric down-conversion driven
by a strong pump pulse, where the interaction length is increased
by cavities both for the pump and the down-converted light. Our
work is thus related to experiments on squeezing \cite{squeezing}
and twin beams \cite{twins,smithey}. Polarization entanglement
between the quantum fluctuations around two macroscopic polarized
beams has recently been created experimentally \cite{bowen}.

Here we aim to create entangled pairs of light pulses such that
the polarization of each pulse is completely undetermined, but the
polarizations of the two pulses are always anti-correlated. Such a
state is the polarization equivalent of an approximate singlet
state of two very large spins. It is thus a dramatic manifestation
of multi-photon entanglement. Starting from a spontaneous process,
the proposed setup builds up entangled states which have very
large photon populations per mode, corresponding to strong
stimulated emission, and thus deserves the name of an
"entanglement laser".

The basic principle of stimulated entanglement creation was
experimentally demonstrated in the few-photon regime in Ref.
\cite{lamas}. To analyze whether the creation of large photon
number entanglement is possible in practice, it is essential to
understand how imperfections in the setup affect the entanglement.
This requires a quantitative measure for the entanglement. We
derive a simple inseparability criterion that is formulated in
terms of the total spin ${\bf J}$ and the total photon number $N$:
if $\l {\bf J}^2 \r/\l N \r$ is smaller than $1/2$, then the state
is entangled. Using this measure we show that strongly entangled
states of very high photon numbers can be generated in the
presence of losses and other imperfections.

\begin{figure}
\center
\includegraphics[width=0.8 \columnwidth]{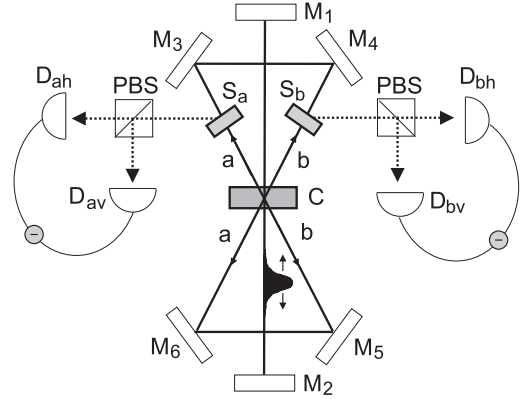}
\caption{Proposed setup for an ``entanglement laser''. An intense
pump pulse propagates back and forth between the mirrors $M_1$ and
$M_2$. Whenever it traverses the non-linear crystal $C$ it creates
polarization entangled photon pairs into the modes $a$ and $b$,
which are counter-propagating pulses inside the cavity formed by
the mirrors $M_3$ to $M_6$. The cavities, which have to be
interferometrically stable, are carefully adjusted such that the
three pulses (pump, $a$ and $b$) always overlap in the crystal.
The fact that $a$ and $b$ propagate in the same cavity
automatically synchronizes the counter-propagating modes. The
number of photons in $a$ and $b$ increases exponentially with the
number of round-trips. They can be switched out of the cavity by
electro-optic switches $S_a$ and $S_b$. The polarization of each
pulse is then analyzed with the help of polarizing beam splitters
(PBS) followed by photo-diodes that give a signal proportional to
the number of photons. Taking the difference between the photon
numbers for the two polarizations behind each PBS corresponds to a
spin measurement. The axis of spin analysis is changed by
appropriate wave-plates in front of the PBS. } \label{setup}
\end{figure}

Let us now study our system in more detail. The source of
entangled light is described by a Hamiltonian \cite{kwiat} \be
H=i\kappa (a^{\dagger }_{h}b^{\dagger }_{v}-a^{\dagger
}_{v}b^{\dagger }_{h})  + h.c. \label{hampdc}, \ee where $a$ and
$b$ refer to the two conjugate directions along which the photon
pairs are emitted, as shown in Fig. 1, $h$ and $v$ denote
horizontal and vertical polarization, and $\kappa$ is a coupling
constant whose magnitude depends on the nonlinear coefficient of
the crystal and on the intensity of the pump pulse. The
Hamiltonian describes two phase coherent twin beam sources,
corresponding to the pairs of modes $a_h, b_v$ and $a_v, b_h$. In
the absence of losses, it produces a state of the form
\begin{eqnarray}
|\psi\rangle = e^{-i \hat{H} t}|0\rangle=
\frac{1}{\cosh^{2}\tau}\sum _{n=0}^{\infty}\sqrt{n+1}
\;\tanh^{n}\tau\;|\psi^{n}_{-}\rangle, \label{pdcstate}
\end{eqnarray}
where $\tau=\kappa t$ is the effective interaction time and
\begin{eqnarray}
&&|\psi^{n}_{-}\rangle = \frac{1}{\sqrt{n+1}}
\frac{1}{n!}(a^{\dagger }_{h}b^{\dagger
}_{v}-a^{\dagger }_{v}b^{\dagger }_{h})^n |0\rangle \nonumber\\
&&= \frac{1}{\sqrt{n+1}}\sum_{m=0}^{n}(-1)^{m}|n\!\!-\!\!m\r_{a_h}
|\,m\r_{a_v}|\,m\r_{b_h}|n\!\!-\!\!m\rangle_{b_v} \,.
\label{psin-}
\end{eqnarray}  All terms in
the expansion in Eq. (\ref{psin-}) have the same magnitude, such
that the observed polarization (the difference in the number of
horizontal and vertical photons) will fluctuate strongly. However,
there is a perfect anti-correlation between the $a$ and $b$
pulses. The state $|\psi\r$ looks the same if the axis of
polarization analysis is rotated by the same amount for the $a$
and $b$ modes. It is the polarization equivalent of a spin singlet
state \cite{durkin}, where the spin components correspond to the
Stokes parameters of polarization, \be
J^A_z&=&\frac{1}{2}(a^{\dagger}_h a_h-a^{\dagger}_v a_v), \,\,\,
J^A_x=\frac{1}{2}(a^{\dagger}_+ a_+-a^{\dagger}_- a_-)
\nonumber\\J^A_y&=&\frac{1}{2}(a^{\dagger}_l a_l-a^{\dagger}_r
a_r), \label{ja}
\ee The spin components can thus be expressed as differences in
photon numbers, where $a_{+,-}=\frac{1}{\sqrt{2}}(a_h \pm a_v)$
correspond to linearly polarized light at $\pm 45^{\circ}$, and
$a_{l,r}=\frac{1}{\sqrt{2}}(a_h \pm i a_v)$ to left- and
righthanded circularly polarized light. The label $A$ refers to
the $a$ modes, cf. Fig. 1. Analogous relations express ${\bf J}^B$
in terms of the $b$ modes. The total spin satisfies $({\bf
J}^A)^2=(J^A_x)^2+(J^A_y)^2+(J^A_z)^2 =
\frac{N_A}{2}\left(\frac{N_A}{2}+1 \right)$. Number states of the
modes $a_h$ and $a_v$ are eigenstates of $J^A_z$ and of $({\bf
J}^A)^2$. The state $|n-k\r_{a_h}|k\r_{a_v}$, has total spin
$j=n/2$ and $J^A_z$ eigenvalue $m=(n-2k)/2$.

The states $|\psi^n_-\r$ of Eq. (\ref{psin-}) are singlet states
of the total angular momentum operator ${\bf J}={\bf J}^A+{\bf
J}^B$ for fixed $j_A=j_B=n/2$. As a consequence, $\l \psi|{\bf
J}^2|\psi\r=0$ also for the state $|\psi\r$ of Eq.
(\ref{pdcstate}). Losses and imperfections lead to non-zero values
for the total angular momentum, corresponding to non-perfect
correlations between the Stokes parameters in the $a$ and $b$
pulses. Since the ideal state of Eq. (\ref{pdcstate}) is highly
entangled, one expects that states in its vicinity are still
entangled. We now present a convenient criterion for entanglement:
for {\it separable} states \be \frac{\l {\bf J}^2 \r}{\l N \r}
\geq \frac{1}{2},\ee where ${\bf J}={\bf J}^A+{\bf J}^B$ and
$N=N_A+N_B$. To prove this, consider $\l {\bf J}^2 \r$ for a
separable state $\rho=\sum_i p_i \rho_i^A \otimes \sigma_i^B$. One
has \be &&\l {\bf J}^2\r=\l ({\bf J}^A)^2 \r +\l
({\bf J}^B)^2 \r + 2 \l {\bf J}^A \cdot {\bf J}^B \r \nonumber\\
&&= \sum_i p_i \l ({\bf J}^A)^2 \r_i + \sum_i p_i \l ({\bf J}^B)^2
\r_i
+ 2 \sum_i p_i \l {\bf J}^A \r_i \l {\bf J}^B \r_i \nonumber\\
&&\geq \sum_i p_i [\l ({\bf J}^A)^2 \r_i+\l ({\bf J}^B)^2
\r_i-2|\l {\bf J}^A \r_i| |\l {\bf J}^B \r_i|] \nonumber\\&& \geq
\sum_i p_i [\l ({\bf J}^A)^2 \r_i+\l ({\bf J}^B)^2 \r_i-2 \alpha_i
\beta_i], \label{ineqs}\ee where $\l {\bf J}^A \r_i=\mbox{Tr}
\rho_i^A {\bf J}^A$, $\l {\bf J}^B \r_i=\mbox{Tr} \sigma_i^B {\bf
J}^B$ etc. Furthermore $\alpha_i=\sqrt{\l ({\bf J}^A)^2
\r_i+\frac{1}{4}}-\frac{1}{2}, \beta_i=\sqrt{\l ({\bf J}^B)^2
\r_i+\frac{1}{4}}-\frac{1}{2}$, and we have used the fact
\cite{inequality} that $|\l {\bf J} \r| \leq \sqrt{\l {\bf J}^2 \r
+ \frac{1}{4} }-\frac{1}{2}$. The last line of Eq. (\ref{ineqs})
can be rewritten as \be &&\sum_i p_i
[\alpha_i^2+\alpha_i+\beta_i^2+\beta_i-2\alpha_i \beta_i]=\sum_i
p_i [(\alpha_i-\beta_i)^2 \nonumber\\ && + \alpha_i + \beta_i]
\geq \sum_i p_i [\alpha_i+\beta_i] \geq \frac{1}{2}(\l N_A \r + \l
N_B \r), \ee where the last inequality follows from $\sqrt{\l {\bf
J}^2 \r + \frac{1}{4} }-\frac{1}{2} \geq \frac{1}{2}\l N \r$,
which is a direct consequence of the relation ${\bf
J}^2=\frac{N}{2}\left(\frac{N}{2}+1 \right)$. Since $N=N_A+N_B$,
this concludes the proof of our criterion. Thus every state that
has $\l {\bf J}^2 \r/\l N \r < \frac{1}{2}$ is entangled. This is
a tight bound. There are separable states that reach $\l {\bf J}^2
\r/\l N \r = \frac{1}{2}$, for example the product state
$|2j\r_{a_h}|0\r_{a_v}|0\r_{b_h}|2j\r_{b_v}$, which in spin
notation corresponds to $|j_A=j,m_A=j\r \otimes |j_B=j, m_B=-j\r$.

It should be emphasized that our criterion is sufficient, but not
necessary. There are entangled states that are not approximate
singlets. Our criterion is specifically designed for the class of
states under consideration and for polarization observables. It
has some similarity to the entanglement criterion for
spin-squeezed states derived in Ref. \cite{sorensen}. The
quantities $\l{\bf J}^2\r$ and $\l N \r$ are simple to calculate,
such that the effects of various imperfections can be studied with
ease. We start by investigating the effect of loss.

Loss in a general mode $c$ corresponds to a transformation $c
\rightarrow \sqrt{\eta}\, c + \sqrt{1-\eta} \, d$, where $d$ is an
empty mode and $\eta$ is the transmission coefficient. Let us
start by assuming that the modes $a_h$ and $a_h$ suffer an equal
amount of loss described by $\eta_A$, while the $b$ modes have a
transmission $\eta_B$. Using Eq. (\ref{ja}) this leads to the
following transformations: \be \l ({\bf J}^{A,B})^2 \r
&\rightarrow& \eta_{A,B}^2 \l ({\bf J}^{A,B})^2 \r + \frac{3}{4}
\eta_{A,B} (1-\eta_{A,B}) \l N_{A,B} \r \nonumber\\  \l {\bf J}^A
\cdot {\bf J}^B \r &\rightarrow& \eta_A \eta_B \l {\bf J}^A \cdot
{\bf J}^B \r. \ee The state before losses, Eq. (\ref{pdcstate}),
has $\l ({\bf J}^A)^2 \r=\l ({\bf J}^B)^2 \r=-\l {\bf J}^A \cdot
{\bf J}^B \r$, $\l N_A^2 \r=\l N_B^2 \r=\l N_A N_B \r$ and $\l N_A
\r=\l N_B \r=\l N \r/2$, which leads to the following expression
for the total angular momentum after losses: \be \l {\bf J}^2 \r
\rightarrow (\Delta \eta)^2 \l ({\bf J}^A)^2 \r
+\frac{3}{8}[\eta_A (1-\eta_A) + \eta_B (1-\eta_B)]\l N \r,
\label{j2l} \ee where $\Delta \eta=\eta_A-\eta_B$. Remembering
that $({\bf J}^A)^2=\frac{N_A}{2}(\frac{N_A}{2}+1)$ one sees that
the first term in Eq. (\ref{j2l}), which depends on $\Delta \eta$,
is of order $\l N \r^2$, while the second term is only $O(\l N
\r)$. If one wants to observe entanglement for large photon
numbers, it is therefore important for the losses (including
detection efficiencies) in the $a$ and $b$ modes to be well
balanced. More precisely, Eq. (\ref{j2l}) together with our
entanglement criterion implies the condition $\Delta \eta \lesssim
\frac{2\sqrt{2}}{\sqrt{\l N \r}}$. An equivalent requirement was
met for $\l N \r$ of order $10^6$ in the experiment of Ref.
\cite{smithey} that demonstrated the strong photon number
correlations of pulsed twin beams by direct integrative detection.
An analogous condition can be derived for a difference in losses
between different polarization modes. If all modes suffer the same
amount of loss, described by a transmission $\eta$, then only the
second term in Eq. (\ref{j2l}) remains, leading to a loss-induced
correction to the ratio $\l {\bf J}^2 \r/\l N \r$ of
$\frac{3(1-\eta)}{4}$, taking into account that the losses also
transform $\l N \r$ into $\eta \l N \r$. This gives a critical
transmission value $\eta_c=1/3$, above which entanglement is
provable by our criterion. The entanglement is thus surprisingly
robust under balanced losses.

So far we have considered a situation where first the ideal state
of Eq. (\ref{pdcstate}) is created, and then it is subjected to
loss. However, in the cavity setup of Fig. 1, which is required to
achieve high photon numbers, photon creation (in the non-linear
crystal) and loss (in the crystal and all other optical elements)
happen effectively simultaneously. It is convenient to transform
to a new basis of modes given by $c_1=\frac{1}{\sqrt{2}}(a_h+b_v),
c_2=\frac{1}{\sqrt{2}}(a_h-b_v), c_3=\frac{1}{\sqrt{2}}(a_v+b_h),
c_4=\frac{1}{\sqrt{2}}(a_v-b_h)$. In this basis the Hamiltonian
(\ref{hampdc}) becomes that of four independent, but
phase-coherent, squeezers, $H=\frac{i\kappa}{2}
\left((c_1^{\dagger})^2-(c_2^{\dagger})^2-(c_3^{\dagger})^2+(c_4^{\dagger})^2+h.c.
\right)$. Introducing the quadrature operators
$x_i=\frac{1}{\sqrt{2}}(c_i + c_i^{\dagger}),
p_i=-\frac{i}{\sqrt{2}}(c_i - c_i^{\dagger})$ gives \be
H=\frac{\kappa}{2} \left( x_1 p_1 - x_2 p_2 - x_3 p_3 + x_4 p_4
\right) + h.c. \label{xp} \ee Writing down the Heisenberg
equations for this Hamiltonian, $\dot{x}_1=i[H,x_1]$ etc., one
sees that $\l p_1^2 \r, \l x_2^2\r, \l x_3^2 \r$ and $\l p_4^2 \r$
become squeezed exponentially, while the fluctuations in the
conjugate quadratures, $\l x_1^2 \r, \l p_2^2 \r, \l p_3^2 \r, \l
x_4^2 \r$ grow correspondingly. In the presence of losses, the
Heisenberg equations have to be replaced by Langevin equations of
the form \be
\dot{x}_1=\kappa(t) x_1 -\lambda x_1 + f_{x1}(t)\nonumber\\
\dot{p}_1=-\kappa(t) p_1 -\lambda p_1 + f_{p1}(t),
\label{dynamics}\ee and corresponding equations for the other
modes. Here the time dependence of $\kappa(t)=\kappa_0 e^{-\Lambda
t}$ takes into account the loss of the pump beam while $\lambda$
is the loss rate of the down-converted light; $f_{x1}(t)$ and
$f_{p1}(t)$ are the quantum noise operators associated with the
losses \cite{scully}, satisfying $\l f_{x1}(t) f_{x1}(t') \r=\l
f_{p1}(t)f_{p1}(t')\r=-i\l f_{x1}(t)f_{p1}(t')\r=\lambda
\delta(t-t')$. Here we have assumed that the loss rate $\lambda$
is the same for all four down-conversion modes $a_h, a_v, b_h,
b_v$. We will discuss the case of unbalanced loss rates below.

Eqs. (\ref{dynamics}) can be integrated explicitly, leading to \be
x_1(t)=e^{\int_0^t k(t') dt'} x_1(0)+\int_0^t dt' e^{\int_{t'}^t
k(t'') dt''} f_{x_1}(t'), \label{xt} \ee  where
$k(t)=\kappa(t)-\lambda$ and $\int \limits_{t'}^t \kappa(t'') dt''
=\frac{\kappa_0}{\Lambda}(e^{-\Lambda t'}-e^{-\Lambda t})$. There
is a corresponding expression for $p_1(t)$ where the sign of
$\kappa (t)$ is flipped.

To understand what these results imply for the polarization
entanglement, one can express the angular momentum in terms of the
quadratures $x_i, p_i$. One finds \be J_z=\frac{1}{2}(x_1 x_2 +
p_1 p_2 - x_3 x_4 - p_3 p_4) \nonumber\\ J_x=\frac{1}{2}(x_1
x_3+p_1 p_3+x_2 x_4+p_2 p_4)\nonumber\\ J_y=\frac{1}{2}(-x_1
p_4+x_4 p_1-x_2 p_3+x_3 p_2). \label{jxp} \ee Introducing the
generic notation $p$ for the quadratures that are squeezed (which
are $p_1,x_2,x_3,p_4$) and $x$ for those whose fluctuations grow
exponentially (which are $x_1,p_2,p_3,x_4$), one sees that all
terms in Eq. (\ref{jxp}) have the generic form $x \cdot p$, and
one finds $\l {\bf J}^2 \r=3(\l x^2 \r \l p^2 \r - \frac{1}{4})$.
The total photon number $N=\frac{1}{2}\sum_i (x_i^2+p_i^2-1)$,
leading to \be \frac{\l {\bf J}^2 \r}{\l N
\r}=\frac{3}{2}\cdot\frac{\l x^2 \r \l p^2 \r - \frac{1}{4}}{\l
x^2 \r + \l p^2 \r -1}. \label{j2xp} \ee

Fig. 2 shows the expected time development of the mean photon
number $\l N \r$ and the ratio $\l {\bf J}^2 \r/\l N \r$ as
determined from Eqs. (\ref{j2xp}) and (\ref{xt}) for realistic
parameter values. The experimentally achievable value for $\kappa$
can be estimated by extrapolating existing experimental results
\cite{lamas} to higher pump laser intensities. A value of
$\tau=\kappa t=1$ for a single pass through a 2mm BBO crystal is
realistic with weakly focussed pump pulses of a few $\mu$J, which
is still below the optical damage threshold. The cavity design of
Fig. 1 including switching elements will have loss rates on the
percent level. Fig. 2 shows that very high photon numbers can be
achieved with just a few round-trips. If balanced losses are the
only imperfection, then the entanglement is very strong even for
large photon numbers, as long as the ``laser'' is far above
threshold, i.e. as long as the rate of creation of entangled
photon pairs is much larger than the loss rate ($\kappa/\lambda
\gg 1$). Note that we are interested in the onset regime, far from
saturation (depletion of the pump).

The photon number $\l N \r$ is limited by the requirement of
observing entanglement in the presence of other imperfections. In
particular, Fig. 2 shows the effect of a difference in the loss
rates between the $a$ and $b$ modes. Suppose that the modes $a_h$
and $a_v$ have one loss rate $\lambda_A$, while $b_h$ and $b_v$
have a different one $\lambda_B$.
Then the quadratures $x_i, p_i$ no longer diagonalize the system.
For example, $x_1$ and $x_2$ satisfy the coupled equations \be
\dot{x}_1=\kappa(t)x_1-\bar{\lambda}x_1-\frac{\Delta \lambda}{2}
x_2+f_{x_1}(t)\nonumber\\ \dot{x}_2=-\kappa(t) x_2-\bar{\lambda}
x_2-\frac{\Delta \lambda}{2}x_1+f_{x_2}(t), \ee where
$\bar{\lambda}=\frac{1}{2}(\lambda_A+\lambda_B),\Delta
\lambda=\lambda_A-\lambda_B$ and $f_{x_1,x_2}$ are the appropriate
noise operators. There are analogous coupled equations for the
pairs $p_1$ and $p_2$, $x_3$ and $x_4$, and $p_3$ and $p_4$. These
equations are diagonal for a new basis of modes $\xi_i,\pi_i$ that
is related to the $x_i,p_i$ by a small rotation, which for $\Delta
\lambda \ll \kappa$ takes the simple form: $x_1=\xi_1+(\Delta
\lambda/4 \kappa) \xi_2, x_2=-(\Delta \lambda/4 \kappa) \xi_1 +
\xi_2, x_3=\xi_3-(\Delta \lambda/4 \kappa) \xi_4, x_4=(\Delta
\lambda/4 \kappa) \xi_3 + \xi_4$, and identical equations for the
$p_i$ in terms of the $\pi_i$. In analogy to the case of balanced
losses, the quadratures $\xi_1,\pi_2,\pi_3$ and $\xi_4$ grow
exponentially, while the quadratures $\pi_1,\xi_2,\xi_3$ and
$\pi_4$ become squeezed. Substituting the above expressions for
the $x_i,p_i$ into Eq. (\ref{jxp}) one finds that, due to the
small rotation between the old and new diagonal modes, the $J_i$
contain terms that are quadratic in the new large quadratures
$(\xi_1,\pi_2,\pi_3,\xi_4)$. This leads to an $O(\l N \r^2)$
contribution to ${\bf J}^2$. The dominating correction to the
ratio $\l {\bf J}^2 \r/\l N \r$ is $\frac{(\Delta \lambda)^2}{32
\kappa^2}\l N \r$, leading to the condition $\frac{\Delta
\lambda}{\kappa} \lesssim \frac{4}{\sqrt{\l N \r}}$ for observing
entanglement. In the regime far above threshold, where $\lambda
\ll \kappa$, this is fairly easy to satisfy even for very large
photon numbers.

The effects of other imperfections can be studied in similar ways.
The most important one is a phase mismatch between the two twin
beams, i.e. a Hamiltonian $H=i\kappa (a^{\dagger }_{h}b^{\dagger
}_{v}-e^{i\phi} a^{\dagger }_{v}b^{\dagger }_{h}) + h.c.$ instead
of Eq. (1). This can be brought to the ideal form by a
transformation $a^{\dagger }_{v}\rightarrow e^{-i\phi/2}
a^{\dagger }_{v}$, $b^{\dagger }_{h}\rightarrow e^{-i\phi/2}
b^{\dagger }_{h}$, which is equivalent to $c_3 \rightarrow
e^{i\phi/2} c_3$, $c_4 \rightarrow e^{i\phi/2} c_4$. This
corresponds to a rotation of the quadratures $x_3 \rightarrow \cos
\frac{\phi}{2} x_3 - \sin \frac{\phi}{2} p_3$, $p_3 \rightarrow
\sin \frac{\phi}{2} x_3 + \cos \frac{\phi}{2} p_3$, and
analogously for $x_4, p_4$.
Similarly to the case of unbalanced losses, this gives a
correction to the ratio $\l {\bf J}^2 \r/\l N \r$ whose dominant
term is $\frac{1}{16} \phi^2 \l N \r$, leading to a condition
$\phi \lesssim \frac{4}{\sqrt{3}\l N \r}$ for observing
entanglement. This means that strong entanglement of a million
photons can be observed if $\phi$ is of order $\pi/1000$. This
level of precision of optical phases is challenging, but
conceivable. Strong entanglement for smaller, but still
considerable, photon numbers is correspondingly easier to achieve.

\begin{figure}
\center \includegraphics[width=\columnwidth]{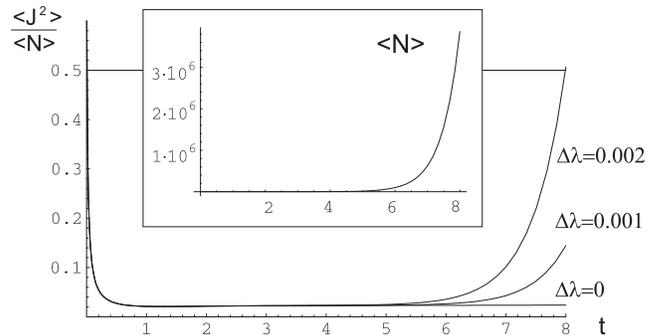}
\caption{Time development of the ratio $\l {\bf J}^2 \r/\l N \r$
and of the mean photon number $\l N \r$. The units are chosen such
that $t=1$ corresponds to a single pass through the crystal. The
initial photon creation rate $\kappa_0=1$, the mean downconverted
photon loss rate $\bar{\lambda}=0.03$ and the pump loss rate
$\Lambda=0.01$. After 8 passes $\l N \r$ reaches the range of
millions. The ratio $\l {\bf J}^2 \r/\l N \r$ is shown for three
different values of the loss rate imbalance $\Delta \lambda$,
namely 0, 0.001 and 0.002. } \label{results}
\end{figure}

An amplitude mismatch in the Hamiltonian, $H=i\kappa (a^{\dagger
}_{h}b^{\dagger }_{v}-f a^{\dagger }_{v}b^{\dagger }_{h}) + h.c.$
with $f$ real, leads to a different degree of squeezing for the
modes $c_1, c_2$ compared to the modes $c_3, c_4$, but not to a
rotation of the quadrature amplitudes, such that the effect on $\l
{\bf J}^2 \r/\l N \r$ does not grow with $\l N \r$.

Another relevant imperfection is a birefringence-related mode
mismatch, corresponding to a Hamiltonian $H=i\kappa (a^{\dagger
}_{h}\tilde{b}^{\dagger }_{v}- \tilde{a}^{\dagger }_{v} b^{\dagger
}_{h}) + h.c.$, where the spatio-temporal modes $\tilde{a}$ and
$\tilde{b}$ of the vertical light differ slightly from the modes
$a$ and $b$ of the horizontal light. In analogy to the case of
losses, one can show that a mode mismatch that affects the $a$ and
$b$ modes in a symmetric way leads to a correction to $\l {\bf
J}^2 \r/\l N \r$ that does not grow with $\l N \r$, which implies
that the birefringence-related walk-off, while important, does not
have to be reduced by orders of magnitude with respect to
experiments on the few-photon level. As before, an asymmetry leads
to an $O(\l N \r)$ effect.  Note that the other major errors that
we have discussed, including the phase mismatch, are also related
to symmetry breaking between the $a$ and $b$ modes. In general,
geometric symmetry between the $a$ and $b$ modes should be
implementable to very high accuracy for the setup of Fig. 1.

In conclusion, the goal of producing strongly entangled
singlet-like states of very large photon numbers seems realistic
with our proposed system. Besides extending the domain where
quantum phenomena have been observed, such states would also have
interesting applications, for example in quantum cryptography
\cite{durkin}. We would like to thank W. Irvine, A. Lamas-Linares
and F. Sciarrino for useful comments. C.S. is supported by a Marie
Curie fellowship of the European Union (HPMF-CT-2001-01205).


\begin{references}
\bibitem{inequality} The inequality $|\l {\bf J} \r| \leq \sqrt{\l {\bf J}^2 \r +
\frac{1}{4} }-\frac{1}{2}$ can be rewritten more intuitively as
$\l {\bf J}^2 \r \geq |\l {\bf J} \r|^2 +|\l{\bf J}\r|$. One can
always rotate the axes of the coordinate system such that only $\l
J_z \r$ is different from zero. The claim is obviously true for a
pure state of fixed total spin $j$. A pure state that is a
superposition of components with different $j$ values is
effectively a mixed state, because off-diagonal terms between
different $j$ do not contribute to $\l {\bf J} \r$ and $\l {\bf
J}^2 \r$. For mixed states $\rho=\sum p_i \rho_i$ we have $\sum
p_i \l {\bf J}^2 \r_i \geq \sum p_i [m_i^2 + m_i]$, where we have
introduced the notation $m_i=|\l {\bf J} \r_i|$. Noting that $|\l
{\bf J} \r_{\rho}|=|\sum p_i \l {\bf J} \r_i| \leq \sum p_i m_i$
and that $\sum p_i m_i^2 \geq (\sum p_i m_i )^2$ one obtains the
desired inequality.
\end{references}
\end{document}